\documentclass[journal,10pt]{IEEEtran}

\usepackage{graphicx,epic,eepic,epsfig,amsmath,latexsym,amssymb,verbatim,subfigure,color}
\usepackage{theorem}

\newtheorem{definition}{Definition}

\newtheorem{lemma}[definition]{Lemma}

\newtheorem{theorem}[definition]{Theorem}

\def\squareforqed{\hbox{\rlap{$\sqcap$}$\sqcup$}}
\def\qed{\ifmmode\squareforqed\else{\unskip\nobreak\hfil
\penalty50\hskip1em\null\nobreak\hfil\squareforqed
\parfillskip=0pt\finalhyphendemerits=0\endgraf}\fi}
\def\endenv{\ifmmode\;\else{\unskip\nobreak\hfil
\penalty50\hskip1em\null\nobreak\hfil\;
\parfillskip=0pt\finalhyphendemerits=0\endgraf}\fi}
\newenvironment{remark}{\noindent \textbf{{Remark~}}}{\qed}

\mathchardef\ordinarycolon\mathcode`\:
\mathcode`\:=\string"8000
\def\vcentcolon{\mathrel{\mathop\ordinarycolon}}
\begingroup \catcode`\:=\active
  \lowercase{\endgroup
  \let :\vcentcolon
  }

\newcommand{\nc}{\newcommand}
\nc{\rnc}{\renewcommand}
\nc{\beq}{\begin{equation}}
\nc{\eeq}{{\end{equation}}}
\nc{\beqa}{\begin{eqnarray}}
\nc{\eeqa}{\end{eqnarray}}
\nc{\lbar}[1]{\overline{#1}}
\nc{\bra}[1]{\langle#1|}
\nc{\ket}[1]{|#1\rangle}
\nc{\ketbra}[2]{|#1\rangle\!\langle#2|}
\nc{\braket}[2]{\langle#1|#2\rangle}
\nc{\proj}[1]{| #1\rangle\!\langle #1 |}
\nc{\avg}[1]{\langle#1\rangle}
\nc{\Rank}{\operatorname{Rank}}
\nc{\smfrac}[2]{\mbox{$\frac{#1}{#2}$}}
\nc{\Tr}{\operatorname{Tr}}
\nc{\tr}{\operatorname{Tr}}
\nc{\id}{\operatorname{id}}
\nc{\1}{\openone}
\nc{\ox}{\otimes}
\nc{\dg}{\dagger}
\nc{\dn}{\downarrow}
\nc{\cA}{{\cal A}}
\nc{\cB}{{\cal B}}
\nc{\cC}{{\cal C}}
\nc{\cD}{{\cal D}}
\nc{\cE}{{\cal E}}
\nc{\cF}{{\cal F}}
\nc{\cG}{{\cal G}}
\nc{\cH}{{\cal H}}
\nc{\cI}{{\cal I}}
\nc{\cJ}{{\cal J}}
\nc{\cK}{{\cal K}}
\nc{\cL}{{\cal L}}
\nc{\cM}{{\cal M}}
\nc{\cN}{{\cal N}}
\nc{\cO}{{\cal O}}
\nc{\cP}{{\cal P}}
\nc{\cR}{{\cal R}}
\nc{\cS}{{\cal S}}
\nc{\cT}{{\cal T}}
\nc{\cX}{{\cal X}}
\nc{\cY}{{\cal Y}}
\nc{\cZ}{{\cal Z}}
\nc{\supp}{{\operatorname{supp}}}
\nc{\var}{\operatorname{var}}
\nc{\rar}{\rightarrow}
\nc{\lrar}{\longrightarrow}
\nc{\polylog}{\operatorname{polylog}}

\def\d{\delta}

\nc{\RR}{{{\mathbb R}}}
\nc{\CC}{{{\mathbb C}}}
\nc{\FF}{{{\mathbb F}}}
\nc{\NN}{{{\mathbb N}}}
\nc{\ZZ}{{{\mathbb Z}}}
\nc{\PP}{{{\mathbb P}}}
\nc{\QQ}{{{\mathbb Q}}}
\nc{\UU}{{{\mathbb U}}}
\nc{\EE}{{{\mathbb E}}}
\nc{\Icoh}{{I^{\rm coh}}}
\nc{\Qca}{{Q_{\rm ss}}}
\nc{\Qcaa}{{Q^{(1)}_{\rm ss}}}
\nc{\Dcaa}{{D^{(1)}_{{\rm ss}\rightarrow}}}
\nc{\Dca}{{D_{{\rm ss}\rightarrow}}}

\nc{\be}{\begin{equation}}
\nc{\ee}{{\end{equation}}}
\nc{\bea}{\begin{eqnarray}}
\nc{\eea}{\end{eqnarray}}
\nc{\<}{\langle}
\rnc{\>}{\rangle}
\nc{\Hom}[2]{\mbox{Hom}(\CC^{#1},\CC^{#2})}
\nc{\rU}{\mbox{U}}

\def\generators{operators\ }

\begin{document}
\title{Codeword Stabilized Quantum Codes}
\author{Andrew Cross, Graeme Smith, John A. Smolin and Bei Zeng
\thanks{Andrew Cross is with the Department of Electrical Engineering, Massachusetts Institute of Technology, Cambridge MA 02139, USA }
\thanks{Andrew Cross, Graeme Smith and John Smolin are with IBM T.J. Watson Research Center, Yorktown Heights, NY 10598, USA}
\thanks{Bei Zeng is  with the Department of Physics, Massachusetts
Institute of Technology, Cambridge, MA 02139, USA} \thanks{GS was partially supported by the UK Engineering and Physical Sciences Research Council, JAS was supported by ARO contract DAAD19-01-C-0056.}}

\date{\today}
\maketitle

\begin{abstract}
We present a unifying approach to quantum error correcting code design
that encompasses additive (stabilizer) codes, as well as all known
examples of nonadditive codes with good parameters.  We use this
framework to generate new codes with superior parameters to any
previously known.  In particular, we find $((10,18,3))$ and $((10,20,3))$
codes.  We also show how to construct encoding circuits for all codes
within our framework.
\end{abstract}

\section{Introduction}
Quantum computers hold the promise of the efficient solution of
problems, such as factoring \cite{Shor94} and simulation of quantum
systems \cite{Feynman82,Feynman86,Lloyd96} that are generally believed
to be intractable on a classical computer.  Furthermore, as the
processor size in state-of-the-art computers continues to scale down
and performance begins to be limited by dissipative effects in logical
processing, it has become increasingly clear that considering
the quantum nature of the components of a classical computer will be
essential in the not-too-distant-future.  In both of these
scenarios---constructing a working quantum computer, or simply
continuing to improve the performance of classical computers---quantum
error correcting codes and ideas from quantum fault-tolerance
\cite{AharonovBenOrACM} will be essential elements in the future
computer engineer's toolbox.

Stabilizer codes are an important class of quantum codes developed in
\cite{Gottesman97,CRSS96}, and are the quantum analogues of classical
additive codes.  An $[n,k]$ stabilizer code encodes $k$ logical qubits
into $n$ physical qubits, and is described by an abelian subgroup,
$S$, of the Pauli group with size $|S| = 2^{n-k}$.  The codespace is
the set of simultaneous eigenvectors of $S$ with eigenvalue $1$.
There is a rich theory of stabilizer codes, and a thorough
understanding of their properties.

Nevertheless, such codes are strictly suboptimal in some
settings---there exist {\em nonadditive codes} which encode a larger
logical space than possible with a stabilizer code of the same length
and capable of tolerating the same number of errors.  There are only a
handful of such examples \cite{RHSS97,SSW07,YCLO07}, and their
constructions have proceeded in an ad hoc fashion, each code working
for seemingly different reasons.  

In the following we present a framework for code design that includes
as special cases stabilizer codes as well as all known nonadditive
codes with good parameters.  We note that the code of \cite{YCLO07}
was presented explicitly in the form we describe below and, indeed,
served as motivation for our studies of the generality of such a
construction.  Our codes are fully described by two objects: a single
stabilizer state $\ket{S}$, and a classical code that generates the
basis vectors of our code from $\ket{S}$.  The stabilizer is chosen
such that it maps all Pauli errors onto only $Z$ errors, though this
may increase their weight.  In this way we map the problem of finding
a quantum code to that of finding a classical code that corrects an
unusual error model.  We have thus unified stabilizer and nonadditive
codes
and rendered both in a form that gives insight into the
classical nature of quantum error-correction.

Our approach is related to the description of nonadditive codes given
in \cite{AC06} in terms of Boolean functions  Our
codeword operators, codeword stabilizer, and effective classical
errors correspond, respectively, to a Boolean function $f$, a matrix
$A_f$, and the ``Zset'' in the language of that work.  
Their approach is essentially dual to ours---in the language we use
here it amounts to first choosing a classical code and trying to
design a stabilizer state whose induced error model is corrected by
the chosen code.  From this perspective, the approach of
\cite{AC06} seems somewhat unnatural, which is perhaps the reason it
has not proved useful for finding new codes.  Both approaches are
closely related to the work of \cite{AKP,GB}.


We describe codes on $n$ qubits that encode $K$ dimensions with
distance $d$ (traditionally written $((n,K,d)))$.  In this framework
we find the original nonadditive $((5,6,2))$ code of \cite{RHSS97} and
the family it generates, the simple family of distance $2$ codes found
in \cite{SSW07}, the $((9,12,3))$ code of \cite{YCLO07}, as well as
new $((10,18,3))$ and $((10,20,3))$ codes.

\section{General construction and Properties}

An $((n,K))$ code will be described by two objects---$S$, a $2^n$
element abelian subgroup of the Pauli group not containing minus the
identity, which we call the {\em word stabilizer}, together with a
family of $K$ $n$-qubit Pauli elements, $W=\{w_l\}_{l=1}^K$, which we
call the {\em word operators}.  There is a unique state $\ket{S}$
stabilized by $S$, {\em i.e.}  $\ket{S}$ satisfies $s\ket{S} =\ket{S}$
for all $s \in S$.  Our code will be spanned by basis vectors of the
form
\begin{equation}
\ket{w_l} \equiv w_l\ket{S}. 
\end{equation}
Since the code vectors should all be different, at most one $w_l$ can
be in $S$.  Typically we will choose $w_1=I$ and later we will prove
this can be done without loss of generality.  Note that $\ket{w_l}$ is
an eigenvector of all $s \in S$ with eigenvalue $\lambda_s=\pm 1$, but
$\ket{w_l}$ is not stabilized by $S$ unless $w_l \in S$.  Each
$\ket{w_l}$ is stabilized by a different stabilizer $w_lSw_l^\dg$.

We would now like to understand the error correction capabilities of
such a {\em codeword stabilized} (CWS) code.  An $((n,K,d))$ code is an
$((n,K))$ code capable of detecting Pauli errors of weight up to
$d-1$, but not $d$, and is said to have minimum distance $d$.  A
distance $d$ code can also be used to correct errors up to weight
$\lfloor (d-1)/2\rfloor$.  The conditions for error correction were
found in \cite{BDSW96,KL97}.  The error correction conditions for a
general code with basis vectors $\ket{w_l}$ are that, in order to
detect errors from a set $\cE$, it is necessary and sufficient to have
\begin{equation}
\bra{c_i}E\ket{c_j} = c_E \d_{ij}
\end{equation}
 for all $E\in \cE$.
For a code of the form described above, this becomes
\begin{equation}
\bra{S}w_i^\dg E w_j \ket{S} = c_E \d_{ij}.
\end{equation}
To correct errors on a fixed number of qubits, 
it is sufficient to study errors of the form $Z^{\mathbf v}X^{\mathbf u}$ with bounded
weight since these form a basis \cite{BDSW96}.
This leads to the {\em necessary and sufficient conditions} for
detecting errors in $\cE$ that for all $E \in \cE$
\begin{equation}
\forall {i\neq j}\ w_i^\dg E w_j \not\in \pm S
\label{detectcondition}
\end{equation}
and 
\begin{eqnarray}
\label{cond1}&&\left( \forall i\ w_i^\dg E w_i \not\in \pm S \right){\rm \ or \ } \\
\label{cond2}&&\left(\forall i\ w_i^\dg E w_i \in S \right) {\rm \ or \ }\\
\label{cond3}&&\left(\forall i\ w_i^\dg E w_i \in -S \right)
\end{eqnarray}
Eq. (\ref{detectcondition}) is the condition that two codewords should
not be confused after an error, while the final three conditions
express that each error must either be detected (Eq. (\ref{cond1})),
or the code must be ``immune'' to it--{\em i.e.} the code is {\em
degenerate}.

\begin{theorem}
\label{theorem:graphstate}
An $((n,K))$ codeword stabilized code with word
\generators $W=\{ w_l\}_{l=1}^K$ and codeword stabilizer $S$  
is locally Clifford-equivalent to a codeword stabilized code with word
\generators $w_l^\prime = Z^{{\mathbf c}_l}$ and codeword stabilizer
$S^\prime$ generated by
\begin{equation}
S^\prime_l = X_l Z^{{\mathbf r}_l}.
\label{graphstabilizer}
\end{equation}
In other words, any CWS code is locally equivalent to a CWS code with
a graph-state stabilizer and word \generators consisting only of $Z$s.
The set of ${\mathbf r}_l$s form the adjacency matrix of the graph.
Moreover, the word \generators can always be chosen to include the
identity.  We call this {\bf standard form}.
\end{theorem}

\begin{proof}
First note that $S$ is local-Clifford equivalent to a graph state due
to \cite{Schlingemann,GKR,VdNDDM}so there is some local-Clifford unitary
$C=\bigotimes_{l=1}^n C_l$ that maps $S$ to $S^\prime$ of the form
(\ref{graphstabilizer}).  In the new basis the word \generators are
$Cw_l C^\dg = \pm Z^{{\mathbf a}_l} X^{{\mathbf b}_l}$, and we have
\begin{equation}
C w_l C^\dag \prod_i \left(S_i^\prime\right)^{({{\mathbf b}_l})_i} = \pm Z^{{\mathbf c}_l},
\end{equation}
so that, letting $w_l^\prime = Z^{{\mathbf c}_l}$, we have
\begin{equation}\nonumber
Z^{{\mathbf c}_l} \ket{S^\prime}  =  \pm Cw_lC^\dg s^\prime \ket{S^\prime}
= \pm Cw_l C^\dg \ket{S^\prime}
 =  \pm Cw_l \ket{S}.
\end{equation}
Since $C$ consists of local Clifford elements, we see that the CWS
code defined by $S^\prime$ and $w^\prime$ is locally Clifford equivalent to the
original code.
 
Finally, to ensure the codeword \generators include the identity we
can choose $\tilde{W}=\{\tilde{w}_l$=$w^\prime_l w^\prime_1\}$ which always has
$\tilde{w}_1={\rm Identity}$.  This can be seen by commuting the
$w^\prime_1$ through the $E$ in the error-correction conditions which
can at worst pick up a sign depending only on $E$.  The two
conditions with $\pm S$ on the right are insensitive to this and
the other two conditions at most change places.
\end{proof}
\begin{figure}
\includegraphics[width=3.in]{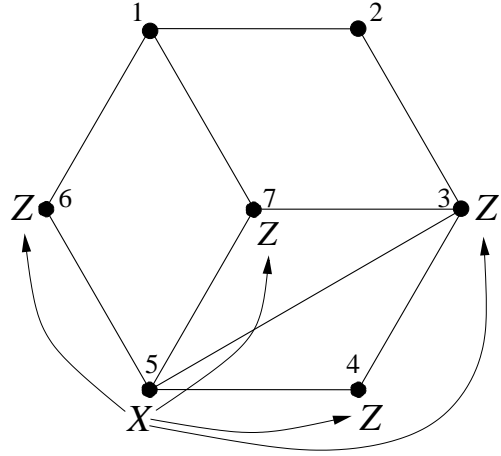}
\caption{Example of the induced error on a graph state:
The state has stabilizer generators $XZIIIZZ$, $ZXZIIII$, $IZXZIIZ$, $IIZXZII$,
$IIZZXZZ$, $ZIIIZXI$, and $ZIZIZIX$.  An $X$ error applied to node 5 in
the lower-left is translated by multiplying with the stabilizer element
$IIZZXZZ$ and turns into $Z$ errors on the nodes indicated.
\label{pushedout}
}
\end{figure}
This structure theorem gives rise to the following lemma, which is
at the heart of our construction:
\begin{lemma}
A single qubit Pauli error $Z$, $X$ or $Y=ZX$ acting on a codeword
$w\ket{S}$ of a CWS code in standard form is equivalent up to a sign to
another (possibly multi-qubit) error consisting only of $Z$s.
\label{lemma}
\end{lemma}
\begin{proof}
Let the error $E_i$ act only on the $i$th qubit. If it
is a $Z$ error the result is immediate.  Otherwise use the fact that
$E_iw\ket{S}=\pm E_i S_i w\ket{S}$, and take $S_i$ to be the generator 
having $X$ on bit $i$.  Then since $E_i=Z_i^{\{0,1\} } X_i$ the 
$X$ in $E_i$ cancels with the $X$ from $S_i$ and we are left with the 
$Z$s from $S_i$ as well as a $Z_i$ if $E_i$ was $Z_i X_i$.
\end{proof}

Lemma \ref{lemma} allows us to construct CWS codes with a satisfying
interpretation: $X$ errors on any qubit are ``pushed'' outwards along
the edges of the graph and transformed into $Z$s.  This is illustrated
in figure \ref{pushedout}.  Similarly $Y$ errors are pushed along the
edges, but also leave a $Z$ behind at their original locations.  Since
all errors become $Z$s, we can think of the error model as classical,
albeit consisting of strange multi-bit errors.  We define this translation
to classical errors by the function $Cl_S(E\in \cE)\rightarrow \{0,1\}^n$:
\begin{equation}
Cl_S(E=\pm Z^{\mathbf v}X^{\mathbf u})={\mathbf v} \oplus \bigoplus_{l=1}^n 
({\mathbf u})_l {\mathbf r}_l
\end{equation}
where ${\mathbf r}_l$ is the $l$th row of the stabilizer's adjacency
matrix (recall from Eq. (\ref{graphstabilizer}) $S_l =X_l Z^{{\mathbf
r}_l}$ defines ${\mathbf r}_l$).
The codeword \generators $w_l=Z^{{\mathbf c}_l}$ will be chosen to so that
the ${\mathbf c}_l$s are a classical code for this error model.

\begin{theorem}
A CWS code in standard form with stabilizer $S$ and codeword
\generators $\{ Z^{\mathbf c}\}_{{\mathbf c} \in {\cal C}}$ detects errors
from $\cE$ if and only if ${\cal C}$ detects errors from $Cl_{S}(\cE)$
and in addition we have for each $E$,
\begin{eqnarray}
\label{clseneq0}Cl_S(E) & \neq&  0\\
\label{complus} {\ \rm or \ } \forall i\ Z^{{\mathbf c}_i}E&=&EZ^{{\mathbf c}_i}\ .
\end{eqnarray}
Thus, any CWS code is completely specified by a graph state stabilizer $S$ and a classical
code $\cal C$.
\end{theorem}
\begin{proof}
When $i \neq j$, $w_i^\dg E w_j \not\in \pm S$ is satisfied exactly
when $Z^{{\mathbf c}_i}EZ^{{\mathbf c}_j} \not \in \pm S$, which is in
turn equivalent to $Z^{{\mathbf c}_i}Z^{Cl_S(E)} Z^{{\mathbf c}_j}
\not\in \pm S$.  In standard form, the only element of $S$ without any
$X$ is the identity, so that this is satisfied exactly when ${\mathbf
c}_i \oplus Cl_S(E) \neq {\mathbf c}_j$.
This is explicitly the classical error-detection condition.

Similarly, when $i=j$, we must satisfy Eqs.~(\ref{cond1}),
(\ref{cond2}) and (\ref{cond3}), whose three possibilities translate directly
to
\begin{eqnarray}
\label{notinS}\forall {\mathbf c} \ Z^{\mathbf c} E Z^{\mathbf c} \not\in \pm S\\
\label{ins}{\rm or \ } \forall {\mathbf c} \ Z^{\mathbf c} E Z^{\mathbf c} \in S\\
\label{notins}{\rm or \ } \forall {\mathbf c} \  Z^{\mathbf c} E Z^{\mathbf c} \in -S.
\end{eqnarray}
Since $Z^{\mathbf c}=I$ for the ${\mathbf c}=0$ codeword,
Eq. (\ref{notinS}) is equivalent to $E \not \in \pm S$ and therefore to
(\ref{clseneq0}). If (\ref{clseneq0}) (and therefore (\ref{notinS})) is not
satisfied, $E \in \pm S$.
If any $Z^{\mathbf c}$
anticommutes with $E$ we have also $E \in \mp S$.  Since no $s\in S$ is also 
in $-S$ this readily implies the equivalence of (\ref{complus}) to 
(\ref{ins}) and (\ref{notins}).
\end{proof}

\begin{remark}
A classical code expressed in quantum terms would
traditionally comprise computational basis vectors that are
eigenstates of $Z$, and therefore the operators mapping one codeword to
another would be of the form $X^{\mathbf c}$ as these are the only errors that
have any effect.  It then might seem odd that standard form for
CWS codes, the intuition of which is to make everything classical,
would employ word operators and effective errors consisting only of
$Z$s.  This choice is arbitrary (one could exchange
$Z$ and $X$ and nothing in the formalism would be affected) and is
made since the usual form of a graph state stabilizer is to have one
$X$ and some number of $Z$s rather than the reverse.  We hope this
historical accident does not cause too much confusion going forward.
\end{remark}

\subsection{Relation to Stabilizer codes}

The CWS framework includes stabilizer codes, and allows them to
be understood in a new way.  We now show that any stabilizer code
is a CWS code, and give a method for determining if a CWS code is
also a stabilizer code.

\begin{theorem}
An $[n,k]$ stabilizer code with stabilizer generators $S_1,
\dots, S_{n-k}$ and logical operations $\bar{X}_1 \dots \bar{X}_k$ and
$\bar{Z}_1 \dots \bar{Z}_k$, is equivalent to the CWS code defined by
\begin{equation}
S = \left \langle S_1 \dots S_{n-k}, \bar{Z}_1 \dots \bar{Z}_k \right \rangle
\label{stabilizerstabilizer}
\end{equation}
and word \generators
\begin{equation}
w_{\mathbf v} = \bar{X}_{1}^{({\mathbf v})_1}\ox \dots \ox  \bar{X}_{k}^{({\mathbf v})_k}
\label{stabilizerws}
\end{equation}
where $\mathbf v$ is a $k$-bit string.
\end{theorem}
\begin{proof}  To see that this CWS code describes the
original code, note that the stabilizer state associated with $S$ is
$\ket{\bar{0}\dots \bar{0}}$, while the codeword generated by
$W_{\mathbf v}$ acting on $\ket{\bar{0}\dots \bar{0}}$ is
$\ket{({\bar{\mathbf v}})_1 \dots (\bar{{\mathbf v}})_k}$.
\end{proof}


\begin{theorem}
If the word \generators of an $((n,K))$ CWS code are an abelian group $W$ 
(not containing $-I$), then the code is an $[n,k=\log_2 K]$ stabilizer code.
\end{theorem}
\begin{proof}
The stabilizer $S$ of the CWS code is a maximal abelian subgroup of
the Paulis (not containing $-I$) therefore it is isomorphic to the
group $S'=\langle X_1 \ldots X_n \rangle$
and the mapping from $S$ to $S'$ is a Clifford operation $C$ (not
necessarily local). This follows from the definition of the Clifford
group as the automorphisms of the Pauli group.  Because this
automorphism group allows one to achieve any bijective mapping that
preserves commutation relations (see Chapter 4 of \cite{Gottesman97}),
the map can further be chosen to map $W$ to $W'=\langle Z_1 \ldots Z_k
\rangle.$
Here we have made use of the facts that all $w \in W$ anticommute with at 
least one $s \in S$ (which implies $S \cap W = \{I\}$) and that
$S'$ is maximal, which allows us to choose for $W'$ any order $K$ group 
made only of $Z$s we like (since {\em all} products of $X$'s are in
$S'$).  Note this nonlocal Clifford mapping is not the same as the
conversion to $Z$s used in Theorem \ref{theorem:graphstate}.

We can now choose $T'$, $\bar{X}'$ and $\bar{Z}'$ as follows:
\begin{eqnarray}
&&\bar{X}'=W'=\langle Z_1 \ldots Z_k \rangle\\
&&\bar{Z}'=\langle X_1 \ldots X_k \rangle\\
&&T'=\langle X_{k+1} \ldots X_n \rangle
\end{eqnarray}
The inverse Clifford operation $C^\dag$ maps these to our stabilizer code
with stabilizer $T$, and logical operations $\bar{X}=W$ and $\bar{Z}$.

It remains to show this is the same as the CWS code we started with.
$T$ is by construction a subgroup of $S$ ($T'$ is explicitly generated
by a subset of the generators of $S'$) and therefore stabilizes
$\ket{S}$.  $T$ also stabilizes all $\bar{x}\ket{S}$, $\bar{x} \in
\bar{X}$, since $T$ and $\bar{X}$ commute.  Using $\bar{X}=W$ we see
these states are exactly the basis states of the CWS code.
\end{proof}

\section{Examples}
We now give some examples of our construction and 
including all known nonadditive codes with good parameters.

\subsection{The $[5,1,3]$ code}

The celebrated $[5,1,3]$ quantum code \cite{BDSW96,KL97} can be written
as a CWS code using Eqs.~(\ref{stabilizerstabilizer}) and
(\ref{stabilizerws}) but another way of writing it demonstrates the
power of the CWS framework.  Take generators corresponding to a ring
graph:
\begin{equation}
S_i = ZXZII \ \ {\rm and \ cyclic \ shifts}\label{ring}.
\end{equation}
This induces effective errors as follows.  Letting 
$\ket{{\rm R}5}$ be the graph state corresponding to the unique
simultaneous $+1$ eigenvector of these generators, we have
\begin{eqnarray}
\nonumber Z_i\ket{{\rm R}5} & = & Z_i\ket{{\rm R}5}\\
\nonumber X_i \ket{{\rm R}5} & = & Z_{i-1}Z_{i+1}\ket{{\rm R}5}\\
Y_i \ket{{\rm R}5} & = & Z_{i-1}Z_iZ_{i+1}\ket{{\rm R}5},
\label{Eq:EffectiveErrorRing}
\end{eqnarray}
where all additions and subtractions are taken modulo 5.  The corresponding
15 classical errors are:
\begin{equation}
\begin{array}{rlllll}
Z:&10000&01000&00100&00010&00001\\
X:&01001&10100&01010&00101&10010\\
Y:&11001&11100&01110&00111&10011\\
\label{ringclassicalerrors}
\end{array}
\end{equation}
We then must choose $w_l = Z^{{\mathbf c}_l}$ where the ${\mathbf
c}_l$s form a classical code capable of detecting pairs of these
errors.  Since no pair of these errors produces $11111$ the codewords
${\mathbf c}_0=00000$ and ${\mathbf c}_1=11111$ will serve, and
together with the stabilizer (\ref{ring}) completely define the
code. Since the $((5,2,3))$ code is known to be unique we need not
otherwise check that our construction is equivalent to the traditional
presentation of this code.  We note also that for $n\geq 7$ a ring
code with codeword operators $I$ and $\otimes_{l=1}^nZ_l$ gives a
$[n,2,3]$ code.

\subsection{The $((5,6,2))$ code}
The first nonadditive quantum code was found in \cite{RHSS97}, and
encodes a six-dimensional space into five qubits with a minimum
distance of two.  This outperforms the best additive five qubit
distance two code, which can have an encoded dimension of at most
four.  The code was originally found as follows: It was known that the
linear programming upper bound was exactly $6$ for a blocklength $5$
distance $2$ code, and in fact it was possible to completely determine
what the weight enumerator \cite{RainsShadow96} of a code meeting
this bound must be.  The authors of \cite{RHSS97} then performed a
numerical search for such a code, and managed to find one.  The
structure of the resulting code was mysterious, and generating larger
codes in a similar fashion seemed intractable (though
\cite{RainsTwo99} showed how to construct a $((5+2l, 2^{2l+1}3,2))$
code from this code).

As a CWS code the $((5,6,2))$ code of \cite{RHSS97} becomes simple.
We again use the ring stabilizer (\ref{ring}) and will have to detect
the induced errors (\ref{ringclassicalerrors}), but since we are seeking a 
distance-2 code we need only consider single errors rather than pairs.
The classical codewords ${\mathbf c}_l$, $l=0\ldots 5$, are
\begin{equation}
\begin{array}{llllll}
00000&11010&01101&10110&01011&10101
\end{array}
\end{equation}
and the code generated by $\ket{c^{{\rm R}5}}$ and $W_l=Z^{{\rm c}_l}$ is
locally Clifford equivalent to the $((5,6,2))$ code of \cite{RHSS97}.
The $((5+2l,2^{2l+1}3,2))$ codes of \cite{RainsTwo99} are also CWS codes
whose graph state is the union of the ring graph and $l$
Bell pair graphs, and whose classical codewords can be derived straightforwardly
from the $((5,6,2))$ classical codewords.

\subsection{The SSW codes}
A family of distance two codes was found in \cite{SSW07}, which
outperforms the family of \cite{RainsTwo99} for odd blocklengths of
eleven or larger.  The codes were originally described in terms of
their codewords as follows.  If $n = 1 \mod 4$, a basis of our code
consists of vectors of the form
\begin{equation}\label{Eq:SSWwords}
\ket{\mathbf x} + \ket{\bar{\mathbf x}},
\end{equation}  
where $\mathbf x$ ranges over all n-bit vectors of odd weight less
than $(n-1)/2$ and $\bar{\mathbf x}$ is the complement of ${\mathbf
x}$, while if $n=3 \mod 4$, we let $\mathbf x$ range over even
weight vectors of weight less than $(n-1)/2$, leading to an encoded
dimension of $2^{n-2}\left( 1 - \frac{{n-1 \choose (n-1)/2}}{2^{n-1}}\right)$.

We now show that these are actually CWS codes.  Indeed, the codeword
stabilizer of this code will be generated by
\begin{equation}
\left\langle X_1Z_2\dots Z_n, Z_1 X_2, Z_1X_3, \dots,  Z_1X_n \right\rangle ,
\end{equation}
with the corresponding stabilizer state being equivalent to a GHZ
state, $(\ket{0}\ket{+}^{\ox n-1}+\ket{1}\ket{-}^{\ox n-1})/\sqrt{2}$.
The codeword \generators are simply $W_{\mathbf x} = X^{({\mathbf
x})_1}Z^{(({\mathbf x})_2,\dots ({\mathbf x})_n)}$ for each allowed
$\mathbf x$, which can immediately be seen to generate, up to local
unitaries, the same codewords as Eq.~(\ref{Eq:SSWwords}).  Putting the
stabilizer into standard form, we find that the graph state it
describes corresponds to a star graph.

\subsection{The $((9,12,3))$ code}
Like the $((5,6,2))$ code, the codeword stabilizer is of the form
\begin{equation}
S_i=ZXZIIIIII{\rm \ and \ cyclic \ shifts}
\end{equation}
The associated classical code correcting the induced errors is:
\begin{equation}
\begin{array}{llll}
000000000 & 100100100 & 010001100 & 110101000\\
000110001 & 100010101 & 011001010 & 111101110\\
001010011 & 101110111 & 011111111 & 111011011
\end{array}
\end{equation}

\section{New Codes}
\subsection{ Ring codes: $((10,18,3))$}
In light of the excellent performance of ring-stabilizers for CWS
codes---the $((5,6,2))$ and $((9,12,3))$ are both of this form---we
have studied larger blocklength codes based on this stabilizer.  This
leads to a new code that outperforms stabilizer codes for 
blocklength $10$.

The blocklength ten code has a codeword stabilizer generated by
$\left\langle Z_{i-1}X_iZ_{i+1}\right\rangle$ and has $18$ word
\generators of the form $Z^{{\mathbf c}_l}$, with ${\mathbf c}_l$
taken from the list
\begin{equation}
\begin{array}{lll}
0000000000&1101001100&0011001010\\
0000011111&0010001001&1111100000\\
1000111110&1100100101&0101101101\\
0001000110&1010010010&0100110100\\
1001010111&1011010001&0110111000\\
0101110010&1110100011&0111111011.\\
\end{array}
\end{equation}

That this code satisfies the required error correction
conditions can be shown by the straightforward (if tedious) technique
of verifying that the associated classical code corrects the
classical noise model induced by the ring stabilizer. 


\subsection{A $((10,20,3))$ Double Ring Code}
We now consider a CWS code with a codeword stabilizer that is not of
the ring form.  In particular, our stabilizer will correspond to the
double ring, with generators
\begin{equation}
\begin{array}{rlrl}
S_1=&XZIIZZIIII&S_6=&ZIIIIXZIIZ\\
S_2=&ZXZIIIZIII&S_7=&IZIIIZXZII\\
S_3=&IZXZIIIZII&S_8=&IIZIIIZXZI\\
S_4=&IIZXZIIIZI&S_9=&IIIZIIIZXZ\\
S_5=&ZIIZXIIIIZ&S_{10}=&IIIIZZIIZX.
\end{array}
\end{equation}
This leads to a $\ket{S}$ that is a $[10,0,4]$ stabilizer
state. Our classical code $\cal C$ giving the codeword s operators is
\begin{equation}
\nonumber
\begin{array}{llll}
0000000000&1100101101&1100000100&0010010010\\
1001100100&0111011011&1101111110&0010111011\\
1001101111&0111010000&1111000101&1011010100\\
0101100000&1011011111&0101101011&0011000001\\
0000101001&1110010110&0001111010&1110111111.\\
\end{array}
\end{equation}

\section{Encoding Circuits}
Thus far, we have focused on the existence and structure of CWS codes.
We now address a question of fundamental importance: {\em What is the
complexity of encoding a CWS code?}  The answer we find is perhaps the
strongest one could hope for: a CWS code will have an efficient encoding 
circuit as long as there is an efficient encoding circuit for the 
{\em classical } code ${\cal C}$.

We will use the fact \cite{RBB} that a graph state $\ket{S}$ whose graph
has edges $E$ is equal to $\prod_{(j,k) \in E} P_{(j,k)} H^{\otimes
n}\ket{0}^{\otimes n},$ where $P_{(j,k)}$ is the two qubit controlled
phase gate, acting on qubits $j$ and $k$: $P\ket{x}\ket{y} =
(-1)^{xy}\ket{x}\ket{y}$.
\begin{theorem}
Let $S$ and ${\cal C}$ define CWS code ${\cal Q}$, $C$ be
a unitary encoding circuit for the classical code ${\cal C}$, and $Q$
be the unitary mapping $\ket{0}^{\otimes n}$ to $\ket{S}$.  Then,
\begin{equation}
U_{ ({\cal Q},{\cal C})} = QC
\end{equation}
is an encoder for ${\cal Q}$.  In particular, since $Q$
has complexity no more than $n^2$, if $C$ has complexity $f(n)$, the
complexity of our encoder is $\max(n^2,f(n))$.
\end{theorem}

\begin{proof}
The $i$th quantum codeword $\ket{{\mathbf c}_i}$ is given by 
$C\ket{i}$
where ${\mathbf c}_i$ is the $i$th codeword of ${\cal C}$.
So,
\begin{eqnarray}
QC\ket{i} & = & \prod_{(j,k) \in E} P_{(j,k)} H^{\otimes n} X^{{\mathbf c}_i} \ket{0}^{\otimes n}\\
 & = & Z^{{\mathbf c}_i} \prod_{(j,k) \in E} P_{(j,k)} H^{\otimes n}\ket{0}^{\otimes n}\\
 & = & Z^{{\mathbf c}_i} \ket{S}
\end{eqnarray}
\end{proof}

\section{Discussion}
We have presented a new framework for quantum codes and shown how it
encompasses stabilizer codes, elucidates the structure of the known
good nonadditive codes, as well as generates new nonadditive codes with
excellent performance.  It should be noted, however, that there do exist
quantum codes outside of our framework, for example those of \cite{PR04}.

Our codeword stabilized codes are described by two objects: First, the
codeword stabilizer that without loss of generality can be taken to
describe a graph state, and which transforms the quantum errors to be
corrected into effectively classical errors.  And second, a classical
code capable of correcting the induced classical error model.  With a
fixed stabilizer state, finding a quantum code is reduced to finding a
classical code that corrects the (perhaps rather exotic) induced error
model.
We also show that CWS codes include all stabilizer codes.  This new
way of thinking of stabilizer codes may help to find new codes with
good properties.  In fact, this method has since been used
\cite{QCO2007} to systematically categorize all codes of $n \le 8$ and
to find a $((10,24,3))$ code as well as slightly better distance-2
codes.

In a future work we hope to expand our work in several new areas.  We
will give algorithms for finding codes (some of which were employed to
find the new codes presented here) as well as bounds on the
computational complexity of the algorithms.  We also hope to find
more new codes, especially of distance higher than three.

\bibliographystyle{IEEEtran}


\end{document}